\documentclass[journal,twoside,web]{ieeecolor}
\usepackage{jsen}
\usepackage{cite}
\usepackage{amsmath,amssymb,amsfonts}
\usepackage{algorithmic}
\usepackage{graphicx}
\usepackage{textcomp}
\usepackage{url}
\usepackage{wrapfig}
\usepackage{enumerate}

\def\BibTeX{{\rm B\kern-.05em{\sc i\kern-.025em b}\kern-.08em
    T\kern-.1667em\lower.7ex\hbox{E}\kern-.125emX}}
\definecolor{abstractbg}{rgb}{0.89804,0.94510,0.83137}
\begin{document}
\title{Design of a Wearable Parallel Electrical Impedance Imaging System for Healthcare}
\author{Bowen Li, Zekun Chen, Xuefei Chen, Luhao Zhang, Shili Liang
\thanks{(Corresponding author: Shili Liang.)}
\thanks{Bowen Li and  Zekun Chen are with the School of Physics, Northeast Normal University, Changchun 130024, China (e-mail: libowen@nenu.edu.cn) }
\thanks{Shili Liang is also with the School of Physics, Northeast Normal University, Changchun 130024, China (e-mail: lsl@nenu.edu.cn)}}

\IEEEtitleabstractindextext{%
\fcolorbox{abstractbg}{abstractbg}{%
\begin{minipage}{\textwidth}%
\begin{wrapfigure}[13]{r}{2.8in}%
\includegraphics[width=2.6in]{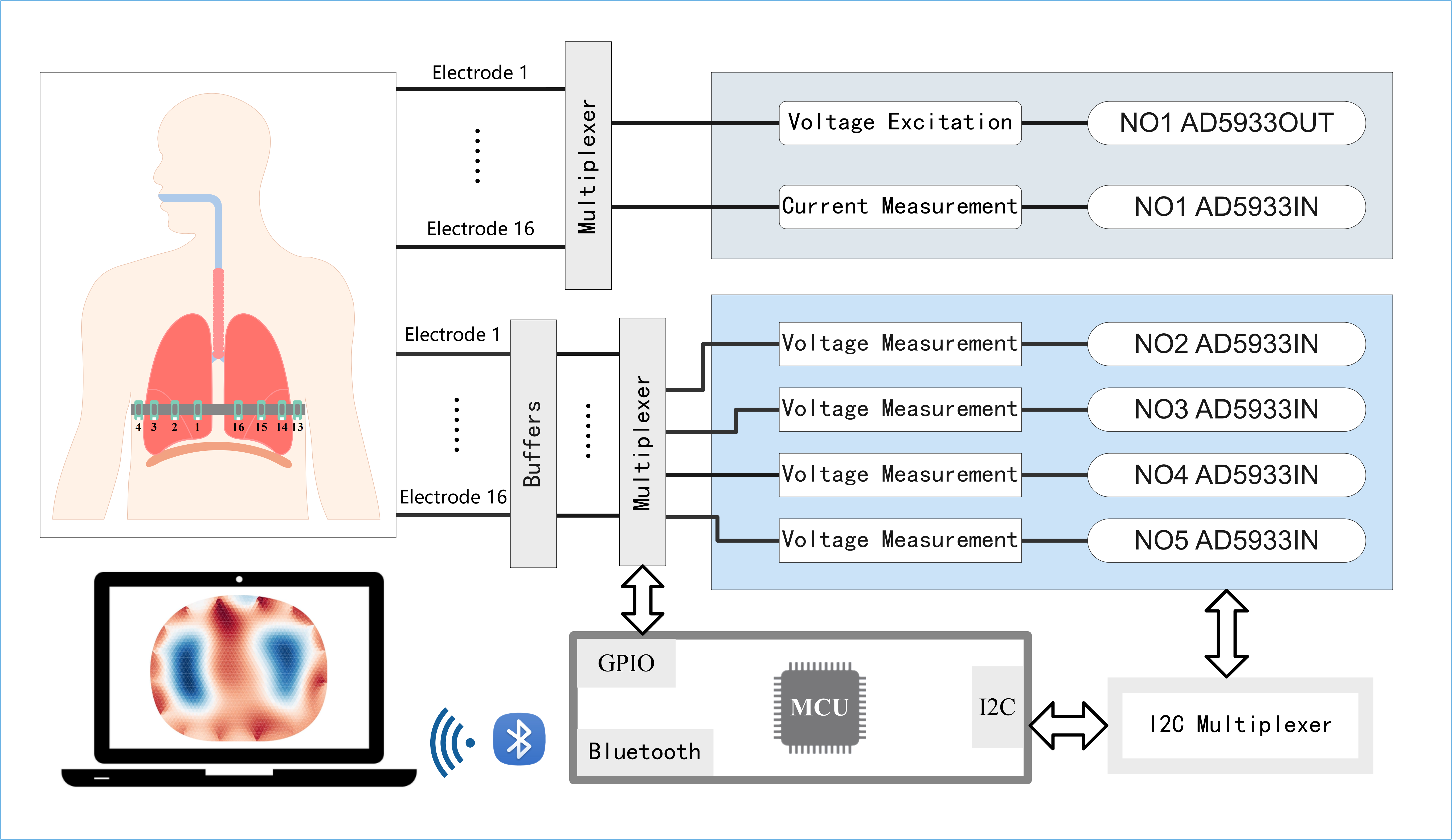}%
\end{wrapfigure}%
\begin{abstract}

A wireless wearable electrical impedance tomography (EIT) system based on the AD5933 chip has been developed, enabling real-time lung respiration imaging. The system adopts a voltage excitation strategy optimized for human impedance characteristics, injecting current through the body by applying a known voltage while measuring the resulting current. Moreover, specific circuit design measures are incorporated to effectively suppress signal oscillations and leakage currents induced by parasitic capacitances. To enhance the data acquisition speed, five AD5933 units are operated in parallel in the system, and several synchronization techniques are adopted to ensure highly synchronized simultaneous measurements. Performance testing shows that the system achieves a SNR exceeding 50 dB, a RSD below 0.3\%, and a RE of less than 0.8\%. Water-tank phantom experiments further validate the system’s favorable spatial resolution, with reconstructed images closely matching the true conductivity distribution. In addition, the system successfully captures the dynamic conductivity variations of the human lungs during respiration, demonstrating its strong imaging capability and practical applicability.
\end{abstract}

\end{minipage}}}

\maketitle

\section{Introduction}
\label{sec:introduction}
\IEEEPARstart{E}{lectrical} impedance tomography (EIT) \cite{lionheart2004eit} is a non-invasive imaging modality that injects low-amplitude, safe alternating currents into the target region via electrode pairs and measures the resulting boundary voltages at the remaining electrodes. The internal impedance distribution is subsequently reconstructed using inverse-problem algorithms.

Boundary voltages can be measured using two primary approaches. In serial systems, the current source selects an electrode pair for current injection, and the voltage measurement module sequentially samples the boundary voltages channel by channel. In parallel systems, the current source also selects an electrode pair for current injection, whereas multiple voltage measurement modules acquire the boundary voltages from multiple channels simultaneously, leading to a substantial improvement in measurement efficiency.

\begin{figure*}[!t]
    \centering
    \includegraphics[width=1 \textwidth]{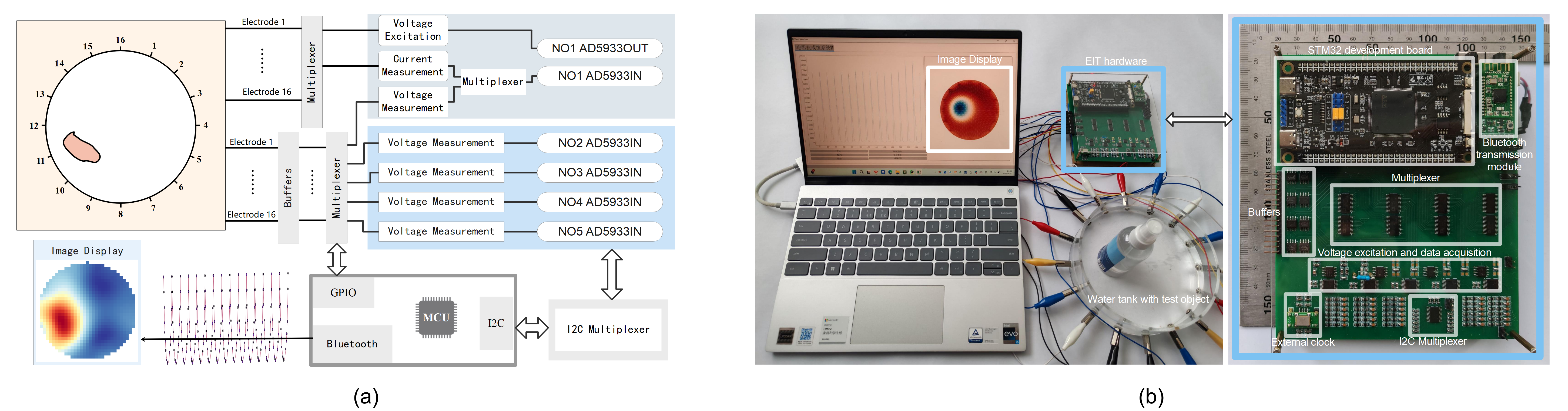}
    \caption{Overview of the EIT system: (a) Block Diagram of Two-Terminal Serial Impedance Measurement; (b) Block Diagram of Four-Terminal Parallel Impedance Measurement; (c) photograph of the system prototype.}
    \label{fig-1}
\end{figure*}

Image reconstruction in EIT is inherently an ill-posed inverse problem \cite{cheney1999electrical}, mainly due to the limited number of measurements compared with the number of unknown pixels. As a result, its spatial resolution is still markedly inferior to that of established imaging modalities such as CT and MRI. Despite this limitation, EIT provides unique advantages, including non-invasive, radiation-free, and real-time imaging capabilities, making it highly attractive for clinical applications.

EIT is widely used in applications such as lung function monitoring \cite{yang2021wireless,hong201510}, brain function assessment \cite{dowrick2015vivo}, cancer detection \cite{choi2007reconstruction}, and human–machine interfaces \cite{zhu2021eit}, with lung monitoring being the most prevalent. By providing real-time visualization of pulmonary gas distribution, EIT assists clinicians in optimizing ventilator parameters and reducing the risk of ventilator-associated lung injury.

Representative commercial EIT systems include Sentec LuMon \cite{sentec_lumon}, Dräger Pulmovista 500 \cite{dräger_pulmovista}, Timpel Enlight 2100 \cite{timpel_ventilation}, Gense \cite{gense_technologies}, and Sciospec’s EIT16 and EIT32/64/128+ platforms \cite{sciospec_eit}. Among these, the Dräger Pulmovista 500 offers real-time visualization of regional lung ventilation, assisting clinicians in evaluating lung collapse, overinflation, and ventilation heterogeneity. Sciospec’s EIT systems commonly use 16 electrodes and support up to 256 electrodes, with real-time frame rates reaching 100 fps and an operating frequency range of 100 Hz–1 MHz.

Recent advances in EIT technology have resulted in the emergence of numerous novel systems, such as the Sheffield Mk3.5 \cite{wilson2001mk3}, ACT-4/ACT-5 \cite{saulnier2007electrical,shishvan2023act5}, OXBACT-3/OXBACT-5 \cite{mcleod1996high,yue2008fpga}, UCLH Mk2/Mk2.5 \cite{romsauerova2006multi,mcewan2006design}, KHU Mark2/Mark2.5 \cite{oh2011fully,wi2013multi,sohal2014electrical}, mfEIT \cite{yang2017multi}, and systems developed at Dartmouth College \cite{hartov2000multichannel,halter2008broadband,halter2004design}. Among these, the Sheffield Mk3.5 system measures impedance at 30 discrete frequencies between 2 kHz and 1.6 MHz, achieving an average SNR of about 40 dB across the entire frequency range. The system uses eight electrodes and consists of eight identical data acquisition boards, each incorporating a DSP to generate excitation signals and perform FFT-based demodulation. The ACT-5 system utilizes 32 electrodes and achieves a frame rate of 27 fps for conductivity imaging. An adaptive current source provides fully programmable current excitation from 5 kHz to 500 kHz, while ECG signals can be displayed in real time during imaging. The KHU Mark2.5 system supports automatic calibration and continuous 24-hour operation at 250 kHz, achieving an SNR above 80 dB. It is implemented on an FPGA platform and includes multiple impedance measurement modules, each equipped with a dedicated constant current source, differential voltage measurement circuitry, and a current source calibrator.

Despite the excellent imaging performance of the aforementioned EIT systems, their high cost hinders widespread deployment. Consequently, there is a strong need to develop a low-cost, easy-to-implement EIT system. In clinical applications, EIT is mainly employed for real-time lung function monitoring. Consequently, the device is required to be compact and portable, while providing rapid data acquisition to accurately track dynamic respiratory changes in the lungs.

This paper presents the development of a portable, low-cost EIT system based on the AD5933 chip. By exploiting the impedance characteristics of the human body, current injection is achieved through voltage excitation and measurement of the resulting current. Unlike traditional current-source–based methods, this approach avoids the need for complete suppression of current-source DC offset. In addition, multiple optimization strategies are adopted to mitigate oscillations and leakage currents induced by parasitic capacitances in the multiplexers, significantly enhancing system stability at high frequencies. Secondly, to enhance data acquisition speed, five AD5933 chips are operated in parallel. I2C bus control is implemented using a TCA9548A multiplexer, while an ICS553 provides a common external clock for all AD5933 chips, thereby ensuring synchronization across multiple acquisition channels. Finally, the PCB layout was optimized for compactness, and Bluetooth wireless communication together with lithium battery power was employed, significantly improving the portability of the device. 

The paper is organized as follows. Section II details the proposed EIT system and its design. Section III reports experimental results from water-tank experiments and a human subject study. Section IV presents a brief conclusion.
\section{Methods}
\subsection{System Architecture}
The impedance acquisition system is implemented using the AD5933, which features an integrated DDS for generating sinusoidal signals from 1 kHz to 100 kHz, along with a 12-bit ADC and an on-chip DSP for precise amplitude and phase measurement.

The photograph of the system prototype with the labeled circuit modules is shown in Fig.\ref{fig-1}(c). The system is composed of an STM32 development board, a power management module, a Bluetooth communication module, a voltage excitation module, a buffering module, a data acquisition module, an I2C multiplexing module, an external clock module, and a signal multiplexing module.

During impedance acquisition, the STM32 controls the first AD5933 through the I2C interface to generate a sinusoidal voltage signal, which is subsequently converted into an excitation voltage by the voltage excitation module. Through multiplexers, the output of the voltage excitation circuit and the input of the current measurement circuit are connected to a selected electrode pair to apply voltage excitation. At the same time, the AD5933 selects between the voltage measurement and current measurement paths to acquire either the boundary voltage or the excitation current passing through the object under test. By controlling the multiplexer channel selection, the STM32 performs a complete scan of the object under test. The AD5933 transfers the processed voltage and current data to the STM32 over the I²C interface, after which the STM32 transmits the computed impedance data to a host computer via Bluetooth for image reconstruction.

The system offers high configurability and can be adapted to different measurement modes as required. It supports both a two-terminal serial measurement mode for basic impedance analysis and a four-terminal parallel measurement mode. The latter separates current injection and voltage sensing paths to mitigate the effects of contact resistance, improve measurement accuracy, and enable simultaneous multi-electrode voltage measurements for faster data acquisition. When operating in the two-terminal serial measurement mode, the impedance is determined by measuring the current produced by a fixed voltage excitation applied to the object under test, as illustrated in Fig.\ref{fig-1}(a).

When operating in the four-terminal parallel measurement mode, the system employs five AD5933 chips in parallel to simultaneously measure the voltages of four electrodes and the excitation current during a single voltage injection. Through repeated channel switching, complete boundary voltage and current measurements under the adjacent excitation pattern are achieved, as illustrated in Fig.\ref{fig-1}(b).

\subsection{Voltage Excitation And Current Measurement}
\subsubsection{Circuit structure analysis}
By applying an excitation voltage to the object under test and measuring the resulting current, the measured current directly represents the excitation current injected into the object. In contrast to fixed-amplitude current excitation methods, this approach eliminates the need for a high-output-impedance current source to stabilize the excitation current and avoids strict linearity requirements between the control signal and the current amplitude. Furthermore, the use of fixed-amplitude voltage excitation eliminates the need to dynamically regulate the output current according to the impedance of the object under test. A further advantage of voltage excitation is that a small DC bias in the excitation signal produces a negligible DC current because of the high DC impedance of the human body, eliminating the need for complete DC bias removal. Moreover, the circuit is simple and robust, operating reliably with a single 3.3 V supply.

\begin{figure}[htbp]
    \centering
    \includegraphics[width=0.7\linewidth]{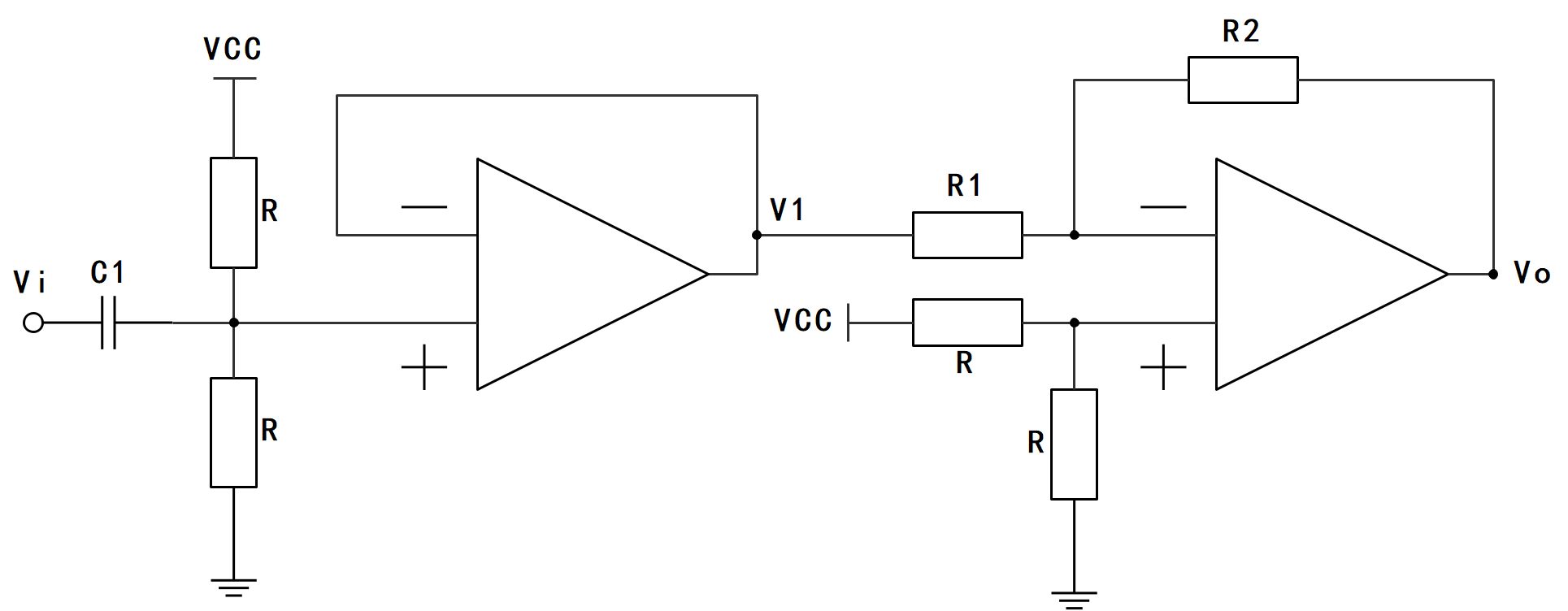}
    \caption{Schematic Diagram of the Voltage Excitation Circuit.}
    \label{fig-2}
\end{figure}

The voltage excitation circuit is shown in Fig.\ref{fig-2}, and the corresponding output voltage is given by:
\begin{equation}
V_0(s) = \frac{V_{cc}}{2} - \frac{R_2}{R_1} \cdot \frac{\frac{R}{2}}{\frac{1}{j \omega C_1} + \frac{R}{2}} V_i(s)
\end{equation}

\begin{figure}[htbp]
    \centering
    \includegraphics[width=0.38\linewidth]{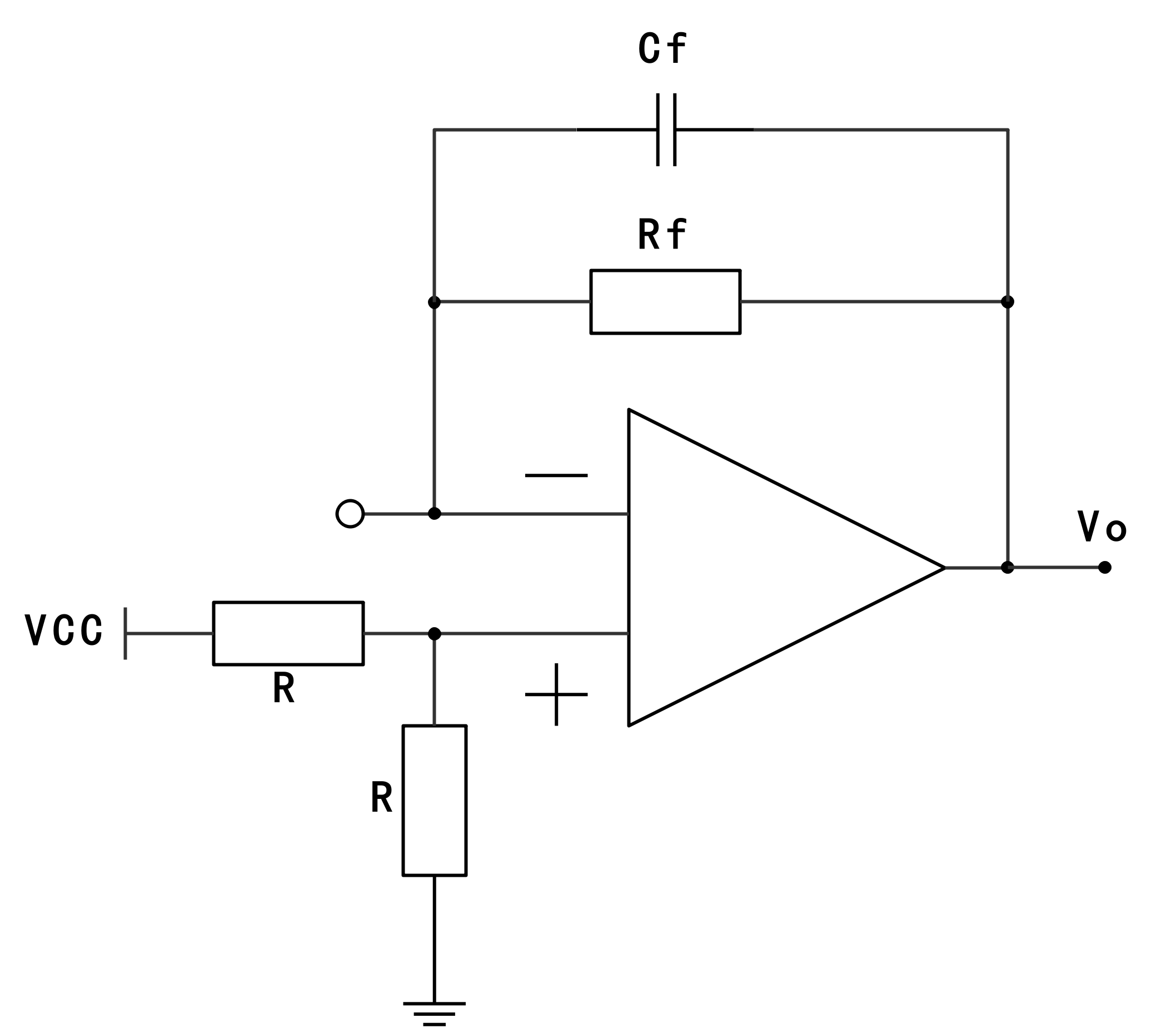}
    \caption{Schematic Diagram of the I–V Conversion Circuit.}
    \label{fig-3}
\end{figure}

The I–V conversion circuit is illustrated in Fig.\ref{fig-3}, and the corresponding output voltage is given by:
\begin{equation}
V_o(s) = - \frac{R_f}{1 + s R_f C_f} \cdot I(s)
\end{equation}
The output of the voltage excitation circuit and the current input of the I–V conversion circuit are connected across the object under test. The voltage excitation circuit generates a 1.98 $V_{p-p}$ AC signal with a DC offset of $\frac{Vcc}{2}$. Because the I–V conversion circuit input is likewise biased at $\frac{Vcc}{2}$, the DC voltage across the object under test is effectively zero, and the applied AC voltage amplitude is 1.98 $V_{p-p}$.

\subsubsection{Circuit stability analysis}
In the impedance acquisition device, parasitic capacitances are unavoidable at the input of the I–V conversion circuit and the output of the voltage excitation circuit as a result of the multiplexers and the physical properties of the PCB. Parasitic capacitance adversely affects circuit stability and increases the likelihood of oscillation. Shen Qidong (1990) systematically analyzed the impact of parasitic capacitance on operational amplifier circuits and proposed effective compensation techniques \cite{Shen1990StrayCapacitanceEffect}, offering valuable guidance for the circuit analysis presented in this work.

\begin{figure}[htbp]
    \centering
    \includegraphics[width=0.36\linewidth]{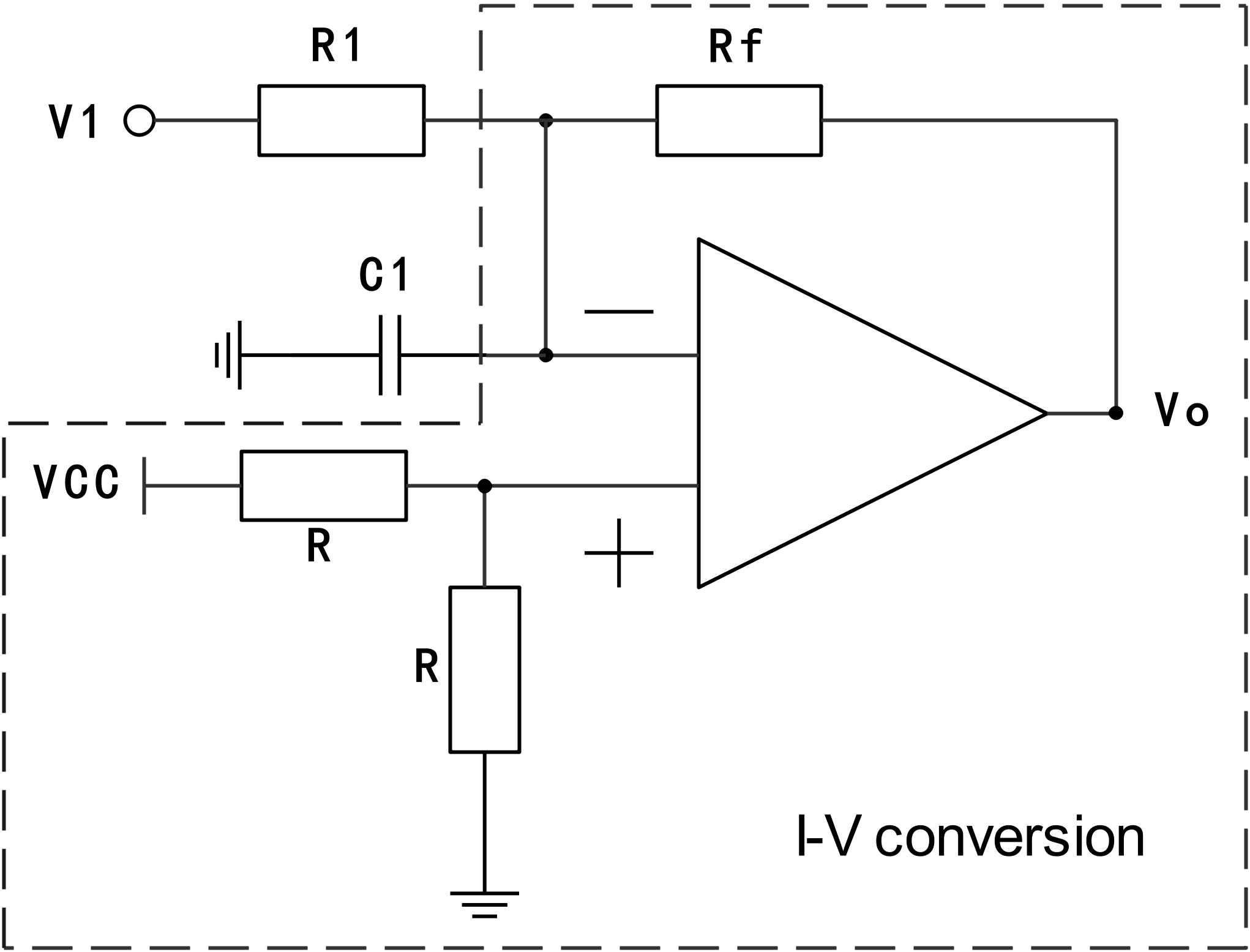}
    \caption{Equivalent Circuit Model of the I–V Conversion Circuit with Parasitic Capacitance.}
    \label{fig-4}
\end{figure}

As illustrated in Fig.\ref{fig-4}, the voltage excitation signal V1 produces a current through resistor R1 that is applied to the I–V conversion circuit. In addition, parasitic capacitance introduced by the multiplexer between the inverting input of the operational amplifier and ground influences the circuit’s frequency response. The loop gain of the circuit can be expressed as:
\begin{equation}
A(s) \cdot F(s) = \frac{A_0}{1 + \frac{s}{\omega_0}} \cdot \frac{R_1}{R_1 + R_f} \frac{1}{1 + \frac{s}{\omega_P}}
\end{equation}
Here, $A_{0}$ and ${\omega_0}$ represent the low-frequency open-loop gain and the dominant pole frequency of the operational amplifier, respectively. Furthermore, $\omega_P = \frac{1}{R_P \cdot C_1}$, $R_p = R_1 \parallel R_f$.

In the feedback circuit, the combination of capacitor $C_{1}$ and resistors $R_{1}$ and $R_{f}$ introduces a lag network. As $C_{1}$ or the feedback resistor $R_{f}$ increases, the resulting second pole shifts to lower frequencies, increasing the likelihood of circuit oscillation. To mitigate this issue, a capacitor $C_{f}$ is typically placed in parallel with the feedback resistor $R_{f}$ as a lead compensation element. The introduced zero cancels the pole when $C_f = \frac{R_1 \cdot C_1}{R_f}$, resulting in improved frequency response and enhanced circuit stability.

\begin{figure}[htbp]
    \centering
    \includegraphics[width=0.45\linewidth]{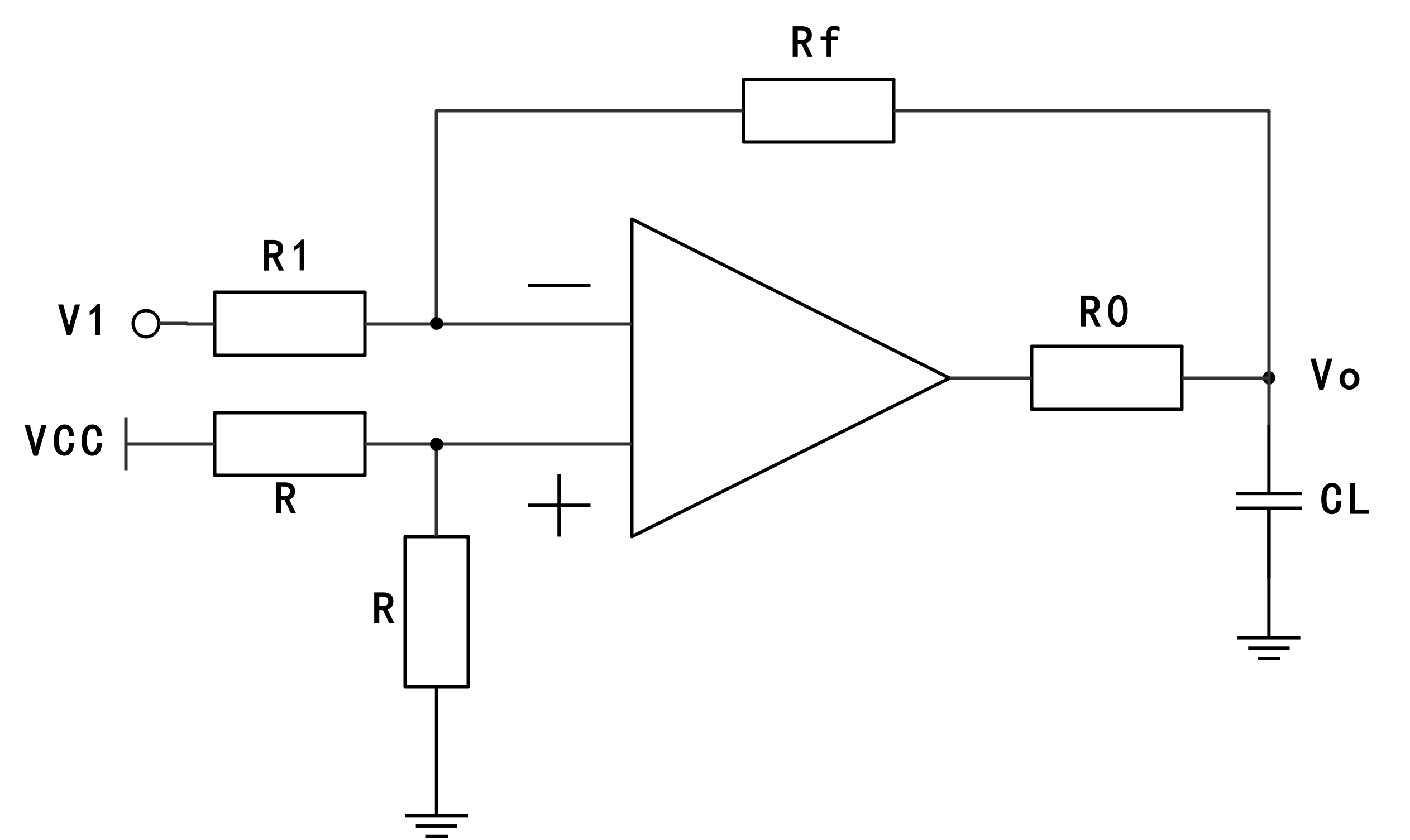}
    \caption{Equivalent Model of the Second Stage of the Voltage Excitation Circuit with Capacitive Load.}
    \label{fig-5}
\end{figure}

Fig.\ref{fig-5} illustrates the second stage of the voltage excitation circuit, where the multiplexer parasitic capacitance acts as a capacitive load $C_{L}$. In combination with the operational amplifier’s internal output resistance $R_{0}$, this introduces a lag network that affects the circuit’s frequency response. The closed-loop gain of the circuit can be expressed as:
\begin{equation}
A_F(s) = \frac{A(s)}{1 + A(s) \cdot F(s)} = \frac{\frac{A_0}{1 + \frac{s}{\omega_0}} \frac{1}{1 + \frac{s}{\omega_Q}}}{1 + \frac{A_0}{1 + \frac{s}{\omega_0}} \frac{1}{1 + \frac{s}{\omega_Q}} \cdot \frac{R_1}{R_1 + R_f}}
\end{equation}
In this expression, ${\omega_0}$ is the dominant pole frequency of the operational amplifier, $\omega_Q = \frac{1}{C_L \cdot R_0}$, where ${\omega_Q}$ corresponds to the second pole frequency arising from the capacitive load and the output resistance of the operational amplifier. The expression for $A_F(s)$ can be rewritten as:
\begin{align}
A_F(s) &= \frac{\frac{R_1 + R_f}{R_1}}{1 + \frac{R_1 + R_f}{R_1 A_0} \frac{s (\omega_0 + \omega_Q)}{\omega_0 \omega_Q} + \frac{R_1 + R_f}{R_1 A_0} \frac{s^2}{\omega_0 \omega_Q}} \notag \\
&= \frac{\frac{R_1 + R_f}{R_1}}{1 - \left( \frac{\omega}{\omega_h} \right)^2 + 2 j \xi \frac{\omega}{\omega_h}} 
\end{align}
The damping factor $\xi$ of the circuit is expressed as $\xi = \frac{\omega_0 + \omega_Q}{2 \omega_h} = \sqrt{\frac{(\omega_0 + \omega_Q)^2}{{4} \cdot \frac{R_1 A_0}{R_1 + R_f} {\omega_0 \omega_Q}}}$, where $\omega_h = \sqrt{\frac{R_1 A_0}{R_1 + R_f} \omega_0 \omega_Q}$. Because $\omega_0 \ll \omega_Q$, $\xi$ can be approximated by $\xi \approx \sqrt{\frac{1}{4 \cdot \frac{R_1 A_0}{R_1 + R_f} w_0 C_L R_0}}$.

As indicated by the damping factor expression, $\xi$ is inversely related to the operational amplifier output resistance $R_{0}$. A lower $R_{0}$ provides stronger damping and improves stability. Although the output impedance of the AD8606 increases with frequency and gain, it remains below 20 $\Omega$ at 100 kHz and a gain of 100, meeting the requirements of the system.

\subsection{Multiplexer and Voltage Measurement}
Channel switching is performed using an ADG706 multiplexer. While this approach may introduce parasitic capacitance relative to traditional relay-based switching, the ADG706 is chosen for its smaller footprint and higher switching speed.

In conventional EIT systems, constant-amplitude current excitation is widely used, where the boundary voltage depends directly on the impedance of the measured region. A programmable gain amplifier (PGA) is therefore typically employed to dynamically scale the voltage so that it remains within the ADC input range. In contrast, this work employs constant-amplitude voltage excitation, eliminating the need for a PGA, as boundary voltage acquisition can be performed solely using a voltage follower circuit.

In the four-terminal parallel measurement mode, directly connecting four voltage measurement channels to four electrodes via multiplexer chips results in each electrode contributing about 200 pF of parasitic capacitance, which arises from the activated multiplexer channels. At frequencies above 50 kHz, parasitic-capacitance-induced leakage currents can alter the internal current field of the object under test and reduce measurement accuracy. To avoid this problem, voltage followers are employed to isolate the 16 inputs of the four multiplexer chips from the electrodes.

\subsection{I2C Multiplexer and External Clock}
An I2C multiplexer, the TCA9548A, is employed to interface with five AD5933 chips. The TCA9548A, provided by Texas Instruments (TI), expands a single I²C bus into eight independently controllable channels. This configuration enables multiple I²C devices with the same address to coexist on a single bus without address conflicts and allows multiple channels to be enabled simultaneously. Through proper register configuration of the TCA9548A, simultaneous command transmission to five AD5933 chips is achieved, guaranteeing synchronized data acquisition.

\begin{figure}[htbp]
    \centering
    \includegraphics[width=0.8\linewidth]{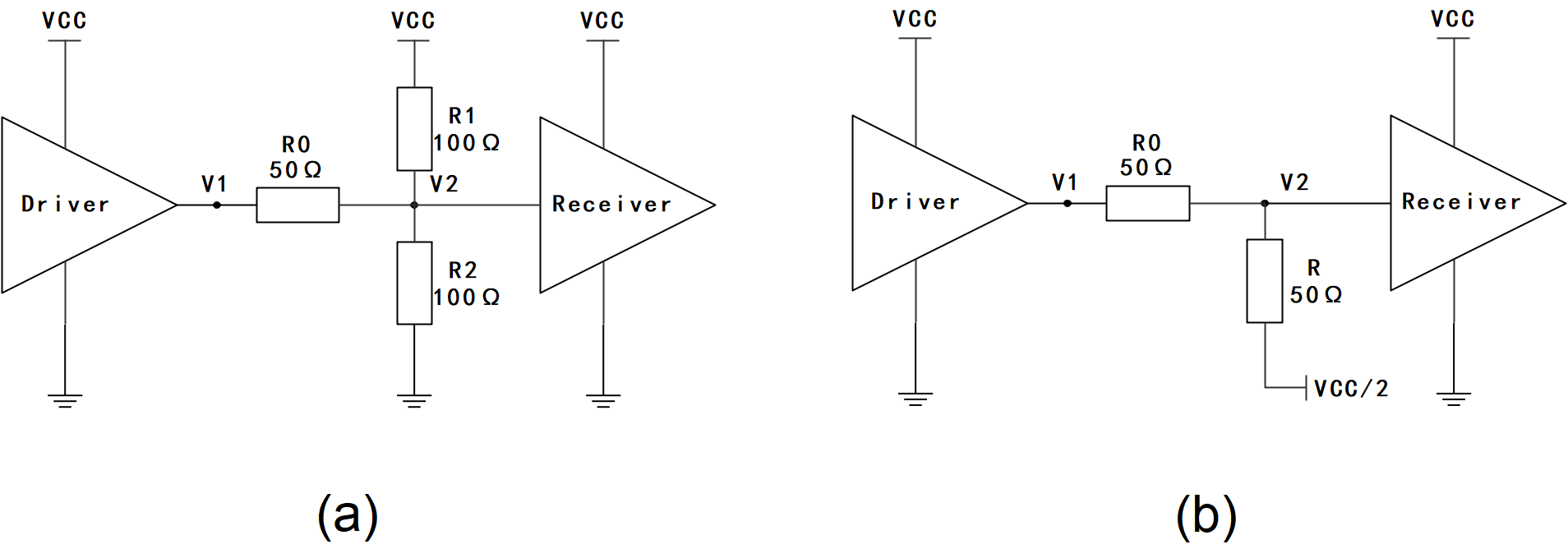}
    \caption{Clock interface circuit between the ICS553 and AD5933 and its equivalent model: (a) clock interface and amplitude-limiting circuit; (b) equivalent model of the amplitude-limiting circuit.}
    \label{fig-6}
\end{figure}

An ICS553 clock buffer from Renesas is employed as the clock distribution device, providing up to four output clock signals derived from a single reference clock input. The use of an external clock is mainly intended to achieve synchronized data acquisition. Since the AD5933 ADC sampling frequency is derived by dividing the external clock frequency by 16, a shared external clock ensures consistent sampling frequencies across the five parallel AD5933 devices. Fig.\ref{fig-6}(a) illustrates the interconnection between the ICS553 clock output and the AD5933 clock input. The circuit is designed to constrain the clock signal amplitude, and the corresponding equivalent circuit is shown in Fig.\ref{fig-6}(b).

\subsection{Microcontroller Program Flow}
\subsubsection{Serial Measurement}
After the system is powered on, the first AD5933 is initialized. A pair of electrodes is then selected from the 16 electrodes through a multiplexer, and a voltage excitation is applied while the corresponding current is measured synchronously. During current measurement, the STM32 initiates a frequency scan by writing a start command to the AD5933 control register and subsequently polls the status register to check DFT completion. Upon completion, the real and imaginary components are read from the AD5933 registers.

Subsequently, the excitation electrode pair is changed through the multiplexer, and the STM32 computes the impedance after each electrode switch. Upon completion of impedance data acquisition, the STM32 sends a complete impedance data frame to the PC for processing. Fig.\ref{fig-8}(a) presents the electrode excitation and measurement sequence for the two-terminal serial mode, and Fig.\ref{fig-9}(a) depicts the corresponding program flowchart.

\subsubsection{Parallel Measurement}
After the system is powered on, the five AD5933 chips are initialized one by one. In four-terminal parallel impedance measurements, a multiplexer selects an electrode pair from the 16 electrodes, and five excitation cycles are performed. For each excitation, the first AD5933 measures the current, while the remaining four AD5933 devices simultaneously acquire voltages from predefined electrodes: (1) electrodes 1, 5, 9, and 13; (2) electrodes 2, 6, 10, and 14; (3) electrodes 3, 7, 11, and 15; (4) electrodes 4, 8, 12, and 16; and (5) electrodes 5, 9, 13, and 1. This configuration ensures that voltage differences between adjacent electrode pairs are measured by the same AD5933 chip, which enhances the common-mode rejection ratio (CMRR) of the system.

\begin{figure}[htbp]
    \centering
    \includegraphics[width=1\linewidth]{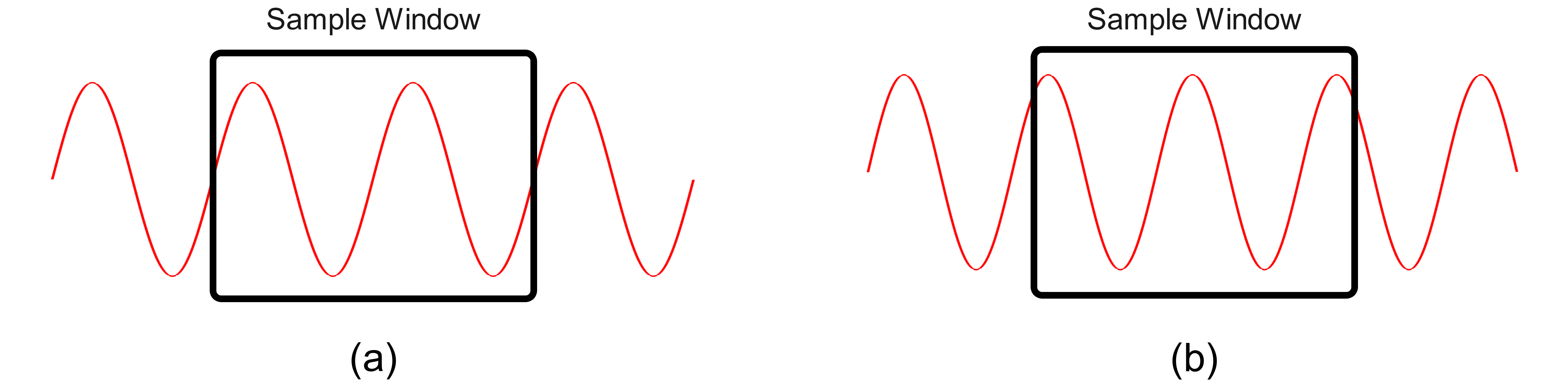}
    \caption{Illustration of the relationship between the DFT sampling window and the excitation signal period: (a) sampling window capturing a non-integer number of excitation cycles; (b) sampling window capturing an integer number of excitation cycles.}
    \label{fig-7}
\end{figure}

During current or voltage acquisition, the first AD5933 is triggered to start a frequency scan for excitation generation and current measurement. Subsequently, the second to fifth AD5933 chips are triggered to start frequency scans for voltage measurement. A voltage excitation signal is applied prior to boundary voltage measurement, with the excitation frequency chosen as an integer multiple of the sampling frequency used for the 1024-point DFT. This configuration guarantees that each 1024-point acquisition window contains an integer number of excitation periods, thus preventing measurement errors arising from waveform truncation. A detailed explanation is shown in Fig.\ref{fig-7}. The sampling frequency for the DFT is expressed as:
\begin{equation}
f_{DFT} = \frac{f_{ADC}}{n}
\end{equation}
where $f_{ADC}$ is the ADC sampling frequency and n is the number of samples acquired (e.g., 1024 samples).

After starting the frequency scan, the STM32 sequentially accesses the five AD5933 chips and polls their status registers to check DFT completion. Upon completion, the real and imaginary components are read from the AD5933 registers.

\begin{figure}[htbp]
    \centering
    \includegraphics[width=1\linewidth]{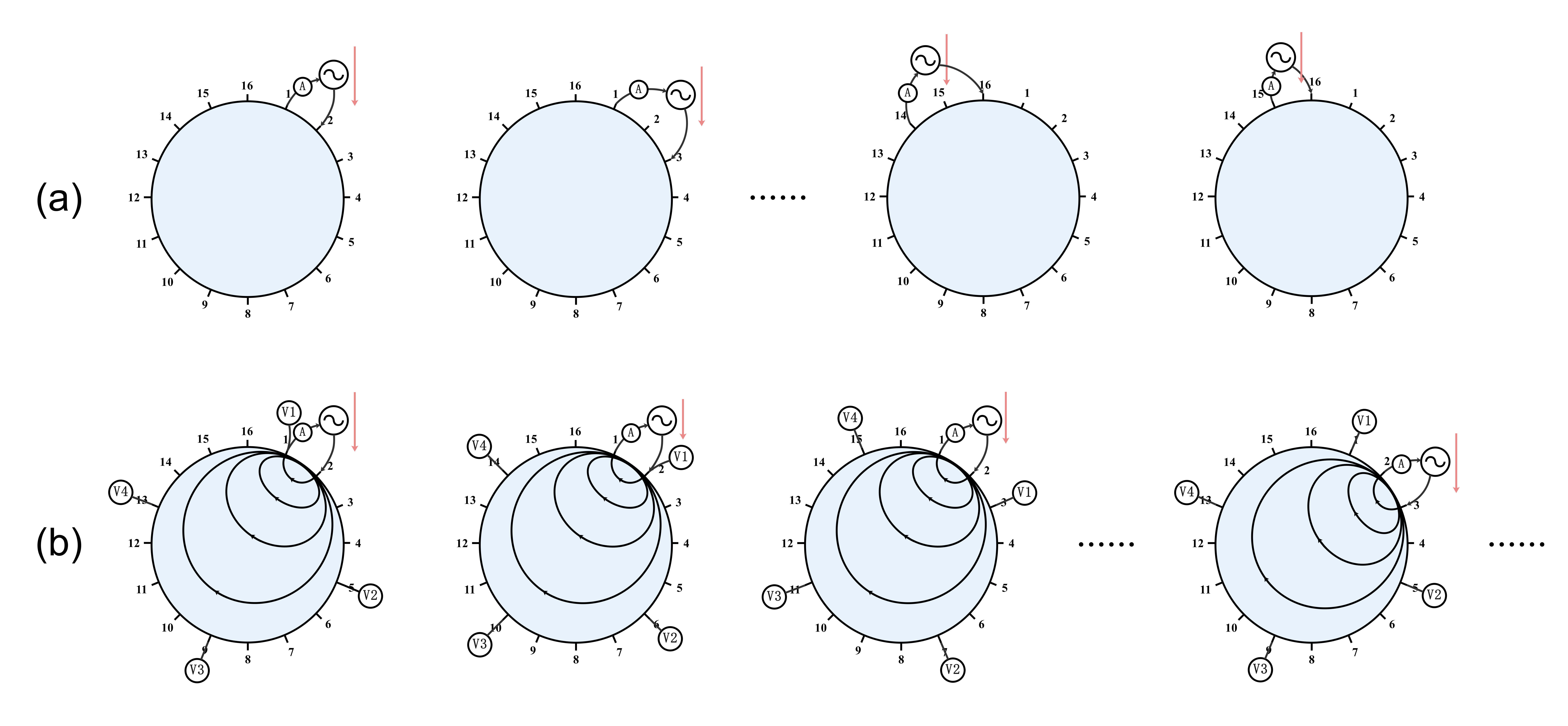}
    \caption{Electrode excitation and measurement sequences under different measurement modes: (a) two-terminal serial measurement mode; (b) four-terminal parallel measurement mode.}
    \label{fig-8}
\end{figure}

\begin{figure}[htbp]
    \centering
    \includegraphics[width=0.9\linewidth]{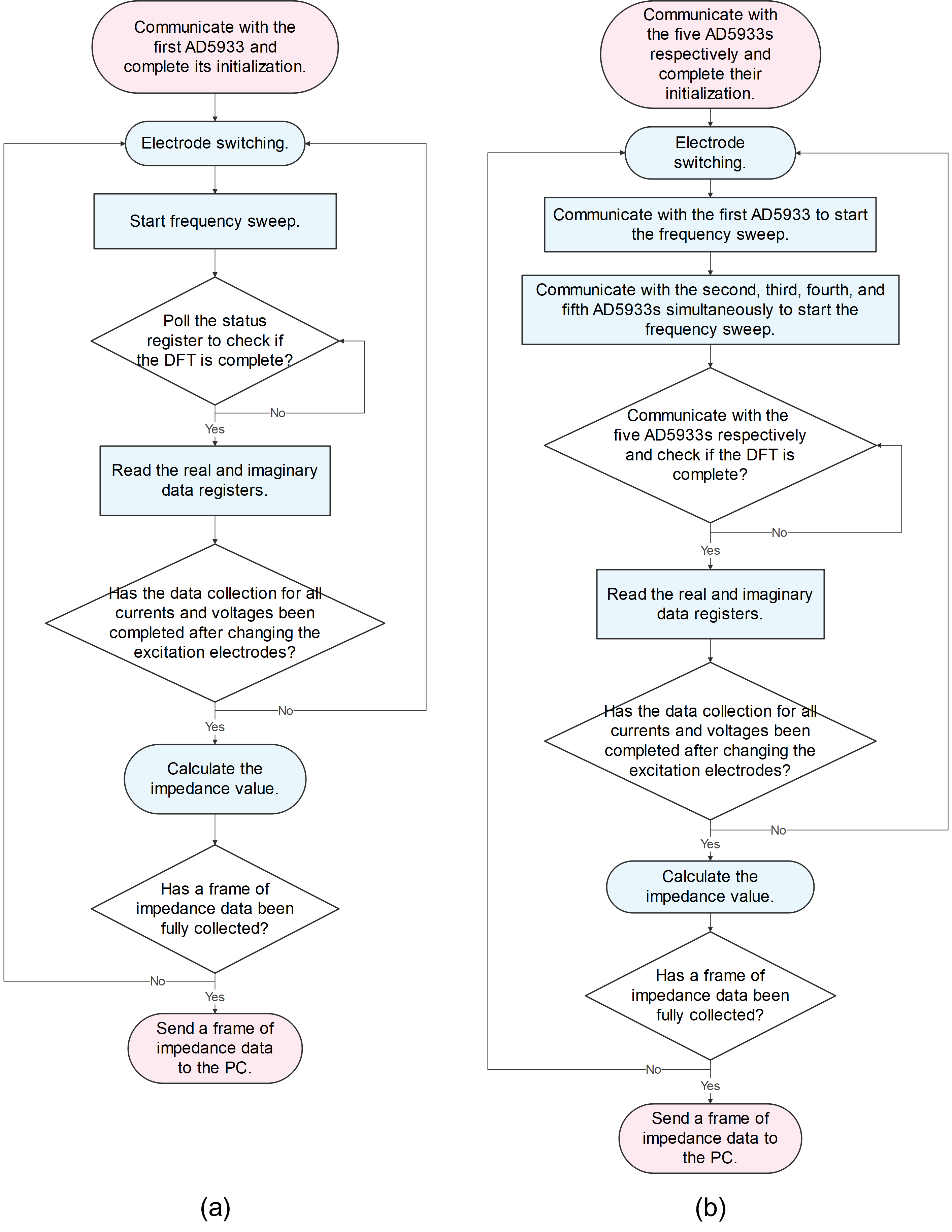}
    \caption{Program flowcharts under different measurement modes: (a) two-terminal serial measurement mode; (b) four-terminal parallel measurement mode.}
    \label{fig-9}
\end{figure}

Subsequently, the excitation electrode pair is changed through the multiplexer, and the STM32 computes the impedance after each switch. Upon completion of data acquisition, a complete impedance data frame is sent to the PC for processing. Fig.\ref{fig-8}(b) shows the excitation and measurement sequence for the four-terminal parallel mode, and Fig.\ref{fig-9}(b) depicts the corresponding program flowchart.

\subsection{Impedance Calculation}
In the two-terminal measurement mode, impedance is determined by measuring the current. With a fixed excitation voltage, the real and imaginary parts of the impedance can be calculated using the measured current and a known gain coefficient.

The phase difference between the current and voltage signals of the object under test is determined through a two-step procedure. First, a pure resistive load is connected to the circuit, and the phase measured by the AD5933 is taken as the system phase. In the second step, the resistive load is replaced by the object under test, and the phase measured by the AD5933 corresponds to the total phase. The phase difference between the current and voltage signals of the object under test is obtained using $\phi_{cal} = \phi_{unk} - \phi_{res}$, where $\phi_{res}$ is the system phase and $\phi_{unk}$ is the measured total phase. Subtracting the two phases removes the system-induced phase offset, enabling accurate determination of the phase difference between the current and voltage signals of the object under test.

Let $|I| \cdot e^{j\phi_{unk}} = Re_{unk} + Im_{unk} j$ and $|I| \cdot e^{j\phi_{res}} = Re_{res} + Im_{res} j$ represent the complex current signals measured by the AD5933 with the object under test and with a pure resistor connected, respectively, where Re and Im denote the real and imaginary components. Based on the phase correction $\phi_{cal} = \phi_{unk} - \phi_{res}$, the complex representation of the calibrated current signal is given by:
\begin{equation}
|I| \cdot e^{j\phi_{cal}} = \left( Re_{unk} + Im_{unk} j \right) \frac{\left( Re_{res} - Im_{res} j \right)}{\sqrt{Re_{res}^2 + Im_{res}^2}}
\end{equation}
The complex impedance Z is given by:
\begin{equation}
Z = \frac{1}{{Gain} \cdot |I| \cdot e^{j\phi_{cal}}} = \frac{1}{{Gain} \cdot \left( Re_{cal} + Im_{cal} j \right)}
\end{equation}
where $Gain$ is a constant gain factor obtained during calibration.

In the four-terminal measurement mode, the voltage signals from two adjacent electrodes, $x_1(n)$ and $x_2(n)$, are processed using a DFT, yielding $X_1(f) = Re_{elec1} + Im_{elec1} j$ and $X_2(f) = Re_{elec2} + Im_{elec2} j$, whose real and imaginary parts correspond to the measured voltage components at the two electrodes. Owing to the linearity of the DFT, the transform of the voltage difference between the two electrodes is given by:
\begin{align}
{DFT}\{x_1(n) - x_2(n)\} &= X_1(f) - X_2(f) \notag \\
&= \left( Re_{\text{elec1}} - Re_{\text{elec2}} \right) \notag \\
&+ \left( Im_{\text{elec1}} - Im_{\text{elec2}} \right) j
\end{align}
Assuming the DFT of the current signal is $X_I(f) = Re_I + Im_I j$, the complex impedance $Z(f)$ is given by:
\begin{equation}
Z(f) = {Gain} \cdot \frac{X_1(f) - X_2(f)}{X_I(f)}
\end{equation}
Likewise, $Gain$ is a constant gain factor obtained during calibration.

\subsection{Graphical User Interface}
PyQt5 provides a Python-based framework for developing graphical user interfaces (GUIs). In this study, it is used to design and implement the GUI. The system software is composed of six modules, including data acquisition, data storage, calibration, imaging, interface mapping, and the main control program. Multithreading is employed to handle real-time data processing and image updates in parallel, ensuring smooth operation of the main user interface.

\begin{figure}[htbp]
    \centering
    \includegraphics[width=0.95\linewidth]{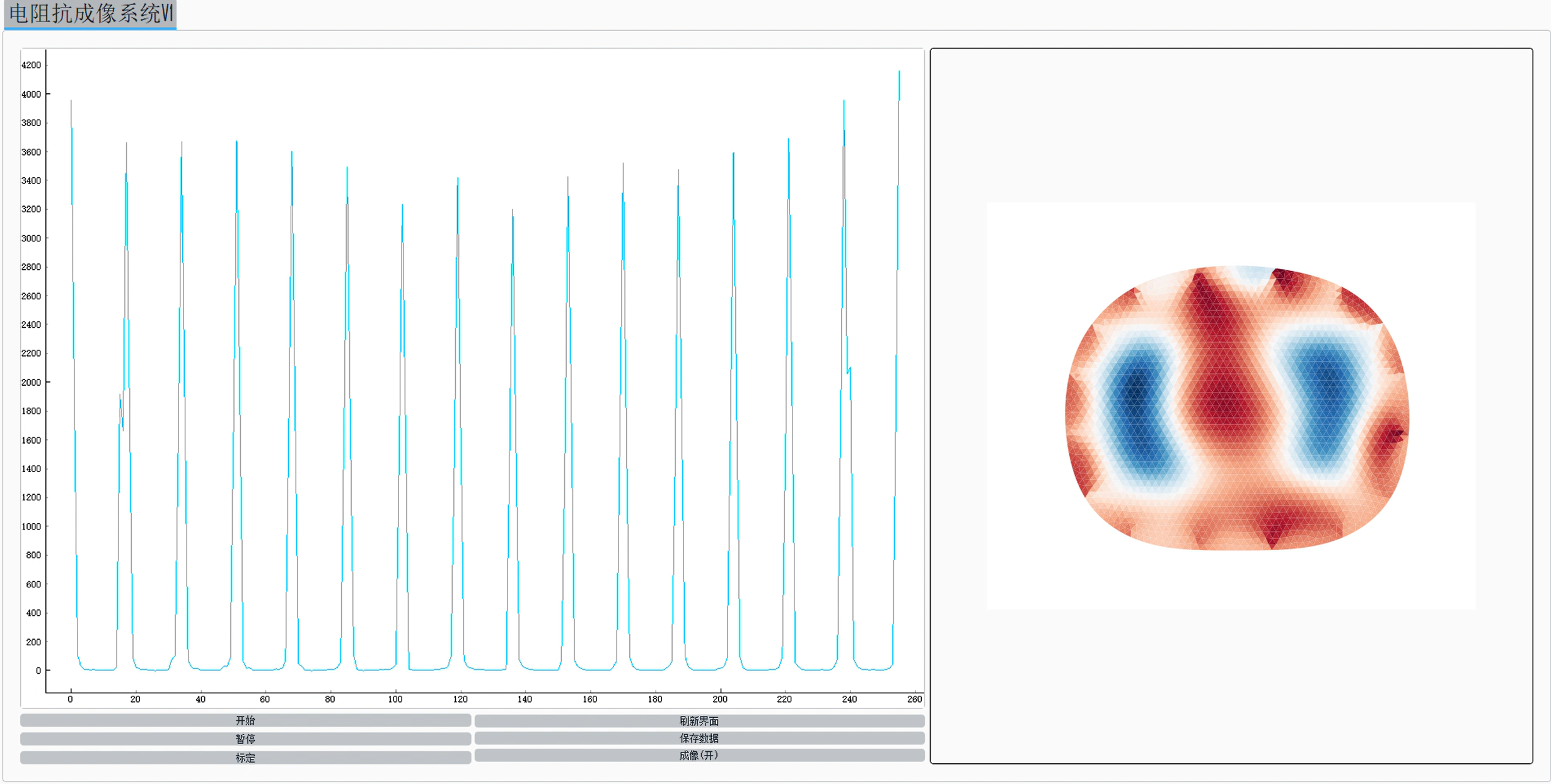}
    \caption{Graphical User Interface (GUI) of the System.}
    \label{fig-10}
\end{figure}

Fig.\ref{fig-10} presents the GUI, which supports real-time imaging, data visualization, data storage, and calibration. Users can intuitively interpret EIT images and easily save the acquired data for offline analysis and experimental reproduction.

The dynamic line plot shown on the left presents the impedance data collected by the acquisition device, where the horizontal axis corresponds to 256 impedance samples and the vertical axis denotes their values. This visualization facilitates intuitive observation of impedance variations and the detection of abnormal patterns. The EIT image displayed on the right uses a color scale to represent impedance values, with red denoting low-impedance regions and blue indicating high-impedance regions. This representation enables clearer observation of the impedance distribution across the scanned region. A set of functional buttons, including “Start,” “Pause,” and “Calibration,” is arranged at the bottom of the interface, enabling users to perform operations such as initiating or stopping data acquisition, refreshing the display, saving data, and controlling image reconstruction. The concise button layout significantly improves ease of use.

\section{Results}
\subsection{System Performance Evaluation}

\begin{figure}[htbp]
    \centering
    \includegraphics[width=0.8\linewidth]{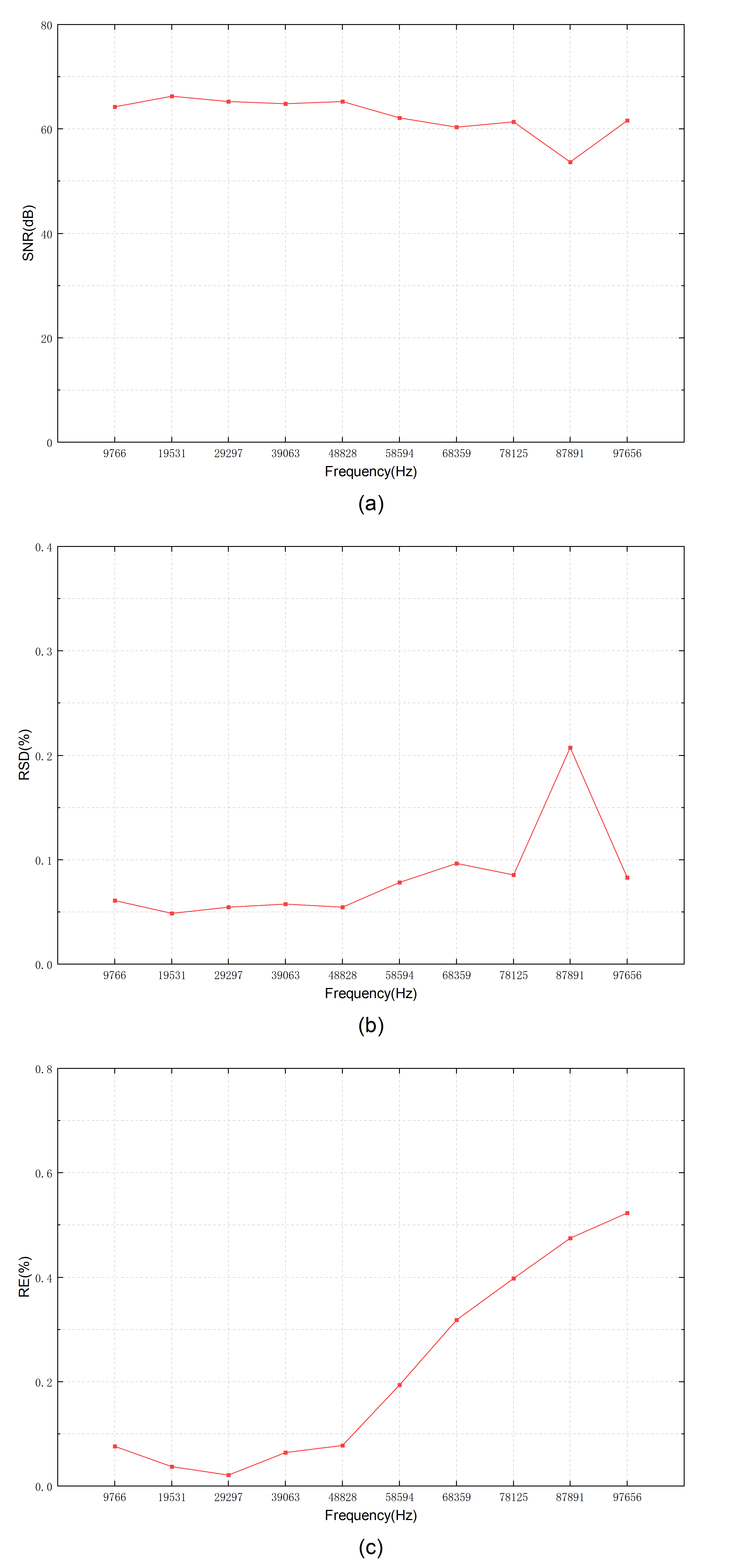}
    \caption{System performance evaluation results: (a) signal-to-noise ratio (SNR) at different frequencies; (b) relative standard deviation (RSD) at different frequencies; (c) reciprocity error (RE) at different frequencies.}
    \label{fig-11}
\end{figure}

To assess the performance of the impedance acquisition device, experiments were conducted using a circular water tank with a 20 cm bottom diameter and a height of 5 cm. Sixteen metal electrodes were evenly arranged around the tank. The tank was filled with water to a depth of approximately 3 cm, and no foreign objects were introduced during the measurements. Impedance measurements were then performed in the four-terminal parallel mode over a frequency range of 8–100 kHz. To reduce frequency leakage, the excitation frequencies were chosen as integer multiples of 1 MHz/1024 when using a 16 MHz crystal oscillator.

At each frequency, 100 measurements were repeated to determine the signal-to-noise ratio (SNR) and the relative standard deviation (RSD). The SNR and RSD were calculated using Eq.\ref{eq：11} and Eq.\ref{eq：12}, respectively.
\begin{equation}
SNR = 10 \log_{10} \left( \frac{\sum_{i=1}^{N} Z_i^2}{\sum_{i=1}^{N} (Z_i - \bar{Z})^2} \right)
\label{eq：11}
\end{equation}
\begin{equation}
RSD = \frac{\sqrt{\frac{1}{N} \sum_{i=1}^{N} (Z_i - \bar{Z})^2}}{\bar{Z}} \times 100\%
\label{eq：12}
\end{equation}
In this context, $Z_i$ represents the value obtained from the i-th measurement, $\bar{Z}$ denotes the average of all measurements, and N is the total number of measurements, which is 100 in this study.

As shown in Fig.\ref{fig-11}(a) and Fig.\ref{fig-11}(b), the SNR and RSD at different frequencies are presented. The SNR ranges from 50 to 70 dB, and the RSD is less than 0.3\% across all frequencies, demonstrating small data fluctuations and good measurement stability.

Subsequently, using the same experimental conditions, the reciprocity error (RE) of the system was measured in the adjacent excitation–adjacent measurement mode, and the RE was calculated using Eq.\ref{eq：13}.
\begin{equation}
RE = \frac{v(c, d) \mid I(a,b) - v(a, b) \mid I(c, d)}{v(c, d) \mid I(a,b)}
\label{eq：13}
\end{equation}
In this context, $v(c, d) \mid I(a,b)$ represents the voltage difference measured across electrodes c and d under current injection between electrodes a and b, while $v(a, b) \mid I(c, d)$ represents the voltage difference measured across electrodes a and b under current injection between electrodes c and d.

As shown in Fig.\ref{fig-11}(c), the reciprocity error (RE) at different frequencies is presented. The RE is less than 0.8\% over the entire frequency range, demonstrating good reciprocity of the system.

\subsection{Water Tank Phantom Imaging Results}
To assess the image reconstruction performance of the system, experiments were also performed in a circular water tank with a diameter of 20 cm and a height of 5 cm. The image reconstruction employed a finite element model with 6,400 triangular elements and was realized using a one-step Gauss–Newton algorithm. All reconstruction procedures were implemented in Python.

\subsubsection{Resolving Capability}
Under the adjacent excitation adjacent measurement mode, detecting small targets in the central region of the water tank is particularly challenging. Therefore, metal targets with diameters of 2.2 mm, 2.9 mm, and 3.6 mm were placed at the center of the tank to evaluate the system’s detection performance. The measurements were carried out in a four-terminal parallel mode at an excitation frequency of 97,656 Hz.

\begin{figure}[htbp]
    \centering
    \includegraphics[width=0.8\linewidth]{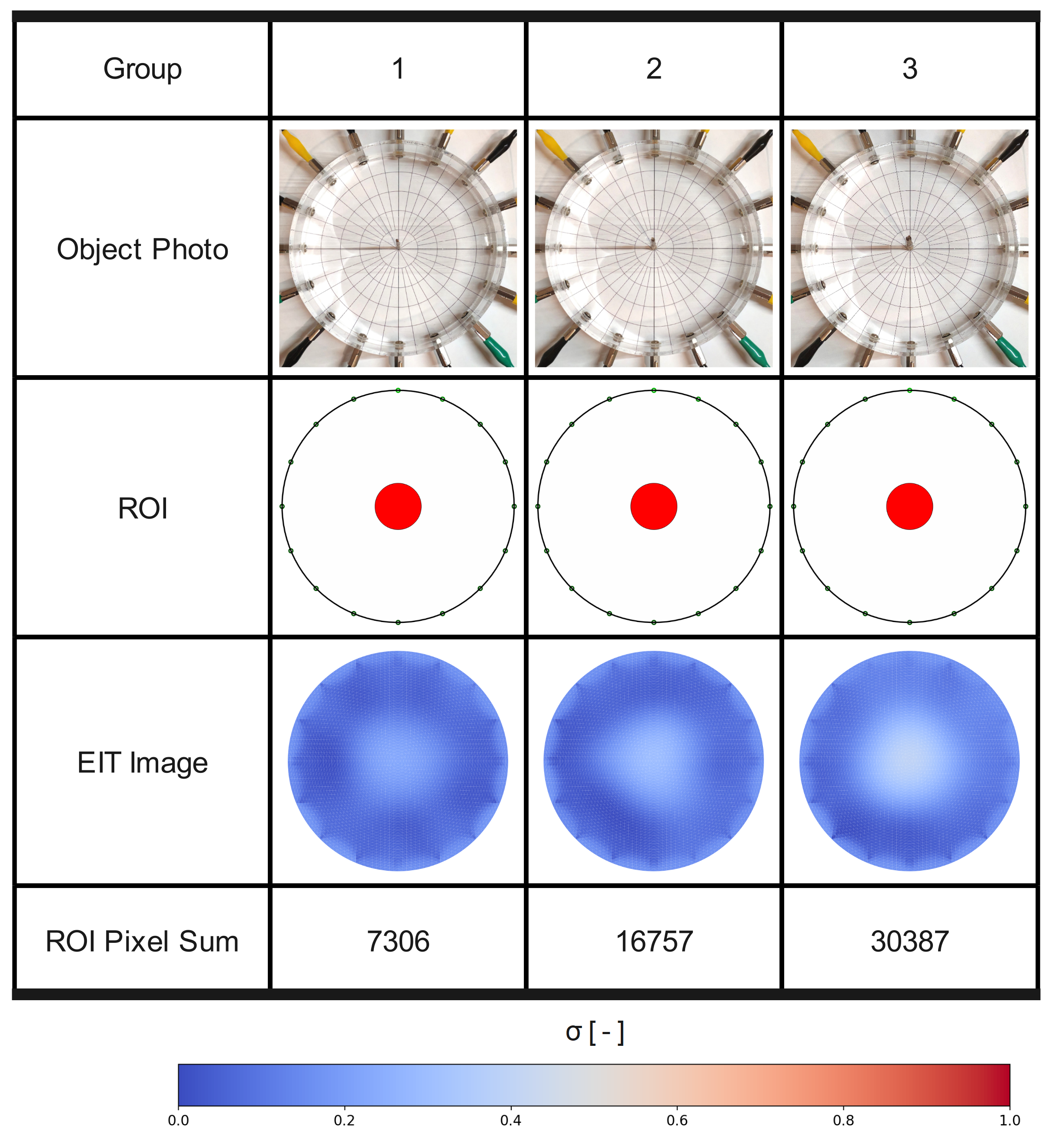}
    \caption{Resolving Capability Evaluation Results for Metal Targets with Different Diameters.}
    \label{fig-12}
\end{figure}

During the experiments, the circular central area of the water-tank electrode model was defined as the region of interest (ROI), and the sum of the finite-element pixel values within this region was computed. As shown in Fig.\ref{fig-12}, photographs of the experimental setup, the reconstructed images, and the corresponding ROI pixel-value sums are presented. The results show a clear increasing trend in the ROI pixel-value sum with increasing metal target diameter, indicating improved resolution for larger metal targets.

\subsubsection{Image Similarity}
To assess the imaging accuracy of the system, experiments were conducted using an insulating cylindrical target placed in the water tank, and the image similarity index (ICC) of the reconstructed images was evaluated according to Eq.\ref{eq：14}. The diameter of the insulating cylinder was 3.8 mm. The measurements were carried out in a four-terminal parallel mode at an excitation frequency of 97,656 Hz.
\begin{equation}
ICC=\frac{\sum_{i=1}^{n}\left(A_i-\bar{A}\right)\left(B_i-\bar{B}\right)}{\sqrt{\sum_{i=1}^{n}\left(A_i-\bar{A}\right)^2}\sqrt{\sum_{i=1}^{n}\left(B_i-\bar{B}\right)^2}}
\label{eq：14}
\end{equation}

In this context, $A_i$ and $B_i$ denote the i-th pixel values of the reconstructed image and the true conductivity distribution image, respectively. $\bar{A}$ and $\bar{B}$ are the corresponding mean pixel values, and n is the total number of pixels.

\begin{figure}[htbp]
    \centering
    \includegraphics[width=1\linewidth]{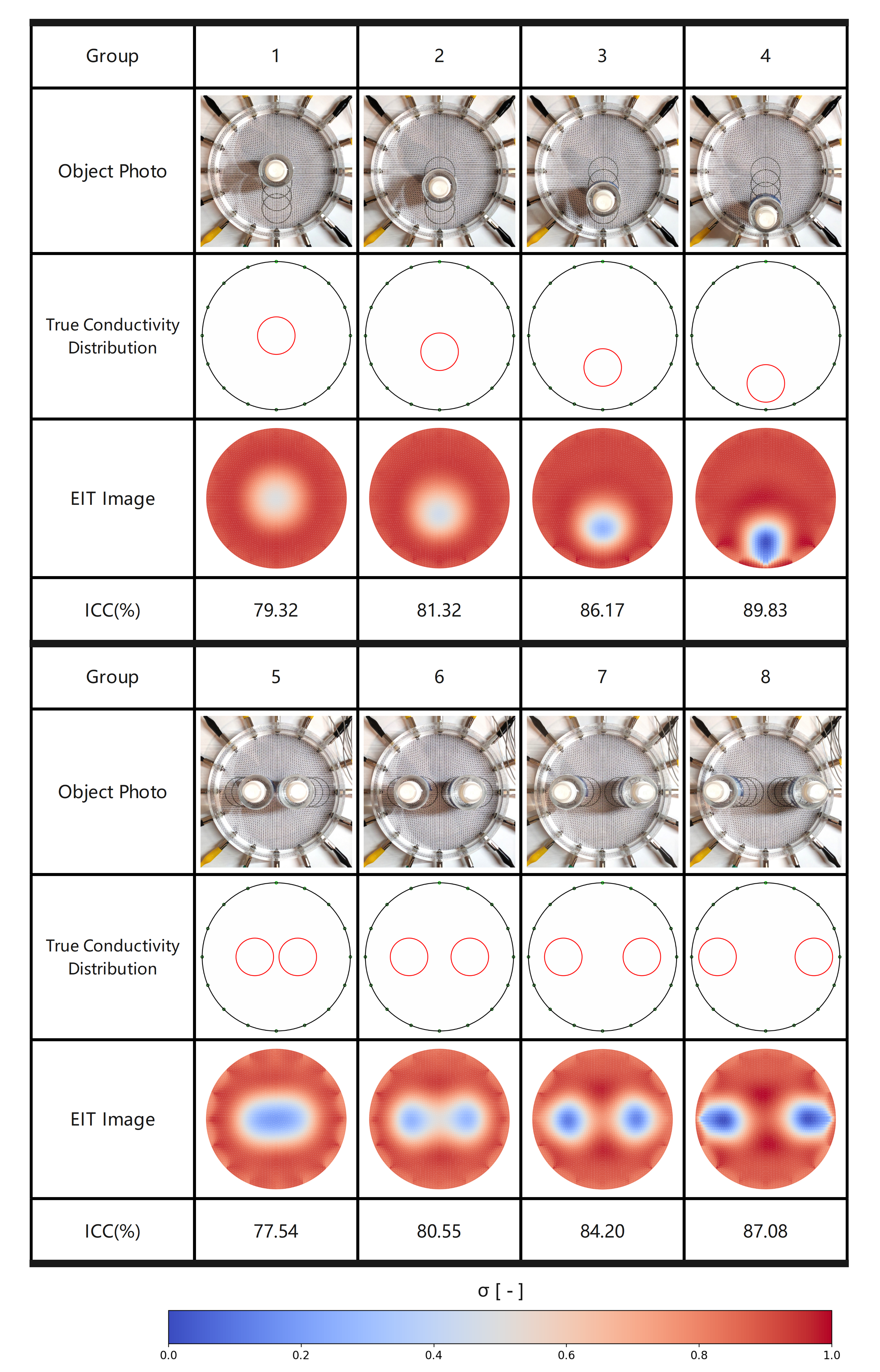}
    \caption{Image Similarity Evaluation Results for Insulating Cylindrical Targets.}
    \label{fig-13}
\end{figure}

As shown in Fig.\ref{fig-13}, the photograph of the experimental setup, the true conductivity distribution image, the reconstructed image, and the corresponding ICC results are presented. The results show that the ICC increases as the insulating cylinder approaches the electrode array, with an average ICC value of 83.25\%.

\subsubsection{Frequency Response}

\begin{figure*}[!t]
    \centering
    \includegraphics[width=0.8\linewidth]{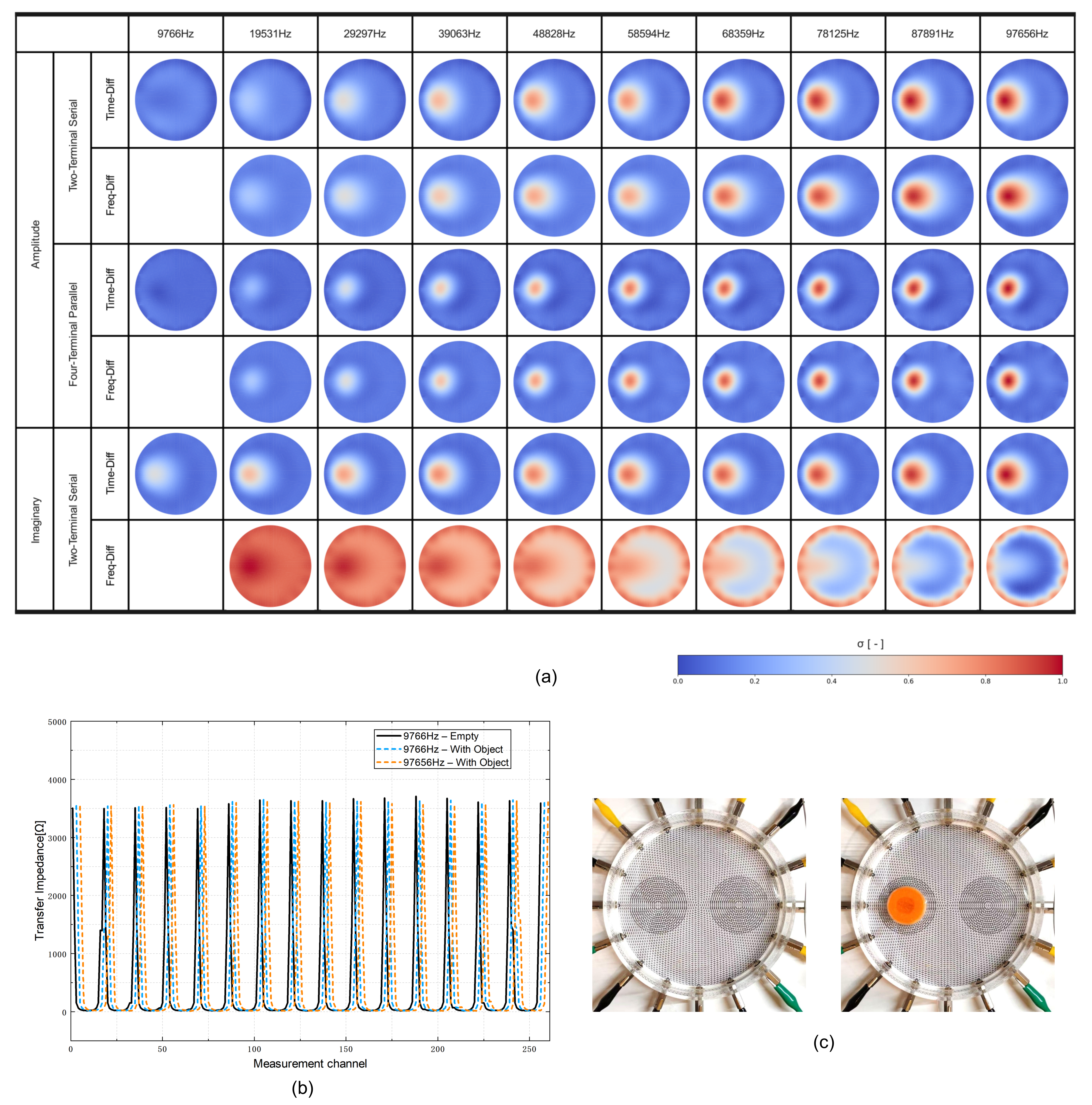}
    \caption{Frequency response evaluation results of the system: (a) image reconstruction results at different frequencies; (b) impedance measurement results at different frequencies; (c) photographs of the experimental setup.}
    \label{fig-14}
\end{figure*}

To assess the imaging performance of the system at different frequencies, experiments were conducted using a carrot with a diameter of 3.3 mm placed in the water tank. Baseline data were first collected with only water in the tank, followed by data acquisition after the carrot was introduced into the tank.

Fig.\ref{fig-14}(a) presents the image reconstruction results obtained using time-difference and frequency-difference imaging under the two-terminal series and four-terminal parallel modes. Fig.\ref{fig-14}(b) shows the four-terminal parallel measurement data acquired at 9,766 Hz without the carrot, at 9,766 Hz with the carrot, and at 97,656 Hz with the carrot. For clarity, the corresponding data curves were shifted rightward by 2 and 4 units, respectively. Fig.\ref{fig-14}(c) shows photographs of the water tank with and without the carrot. The reconstructed images show that the color representation of the carrot changes with the test frequency. This phenomenon is attributed to the increase in electrical conductivity of biological tissues with increasing frequency.

\subsection{Lung Respiratory Imaging Results}

\begin{figure*}[!t]
    \centering
    \includegraphics[width=0.9\linewidth]{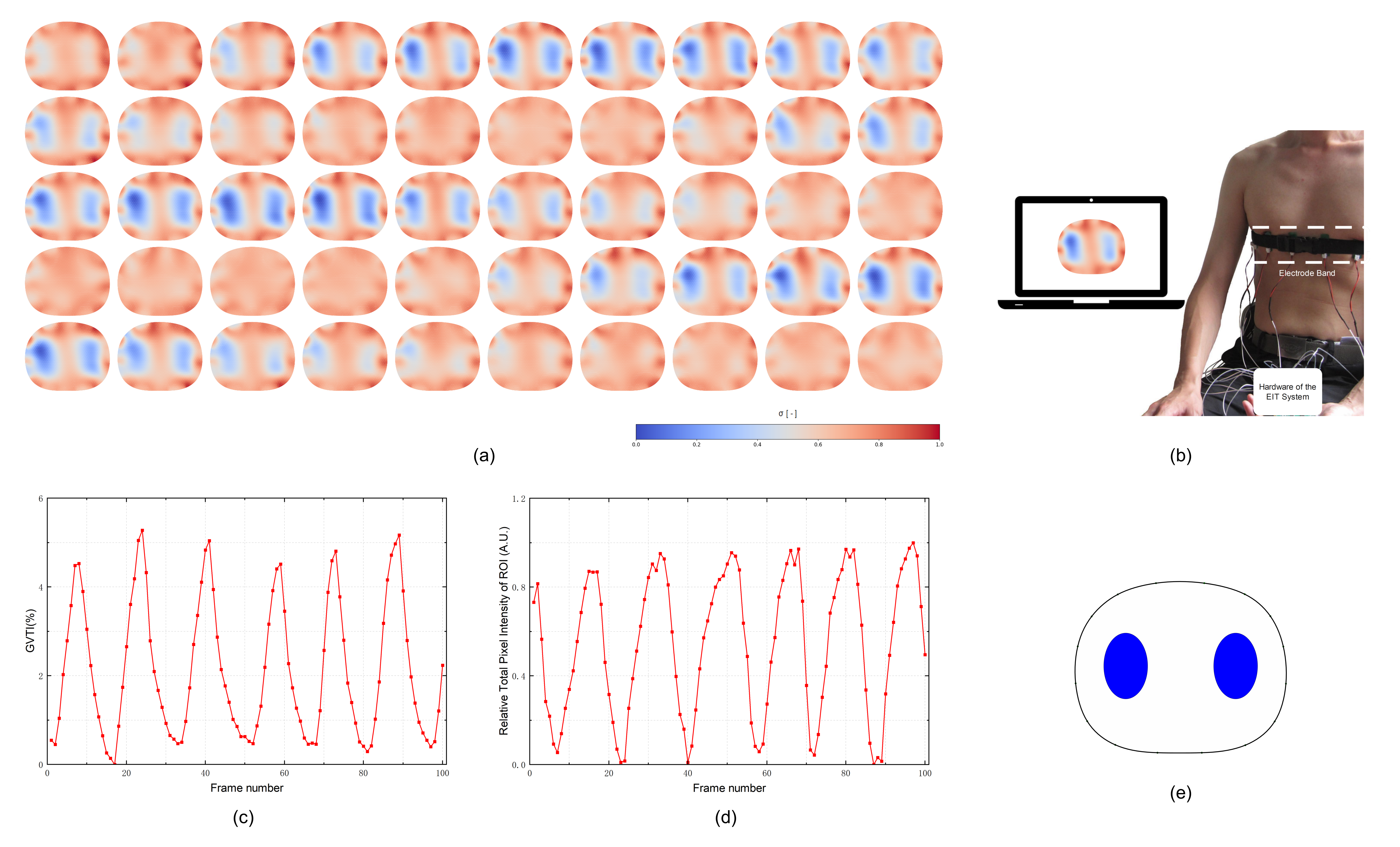}
    \caption{Lung respiratory imaging results: (a) dynamic reconstruction images during respiration; (b) photograph of the electrode arrangement on the thorax; (c) global variation of boundary impedance (GVBI) versus time; (d) summed pixel values within the ROI of the reconstructed conductivity images versus time; (e) ROI region in the lung cross-section.}
    \label{fig-15}
\end{figure*}

For the human lung respiratory dynamic image reconstruction experiment, data were acquired from a 28-year-old male volunteer with a BMI of 21.9. A four-terminal parallel measurement mode was used with an excitation frequency of 97.656 Hz. Time-difference imaging was applied, and the reconstruction was implemented using a one-step Gauss–Newton algorithm.

First, sixteen electrodes were evenly distributed around the subject’s thorax using a PCB-based electrode belt and connected to the impedance acquisition system, as illustrated in Fig.\ref{fig-15}(b). After the subject resumed normal breathing, the system recorded 100 frames of impedance data. Each frame consisted of 256 valid boundary impedance measurements acquired using the adjacent excitation–adjacent measurement pattern. In addition, the global variation of boundary impedance (GVBI) was calculated for each frame according to the following equation:
\begin{equation}
GVBI=\frac{1}{256}\sum_{chn=1}^{256}Z_{chn}
\end{equation}
Here, $Z_{chn}$ represents the boundary impedance value of the chn-th channel.

As shown in Fig.\ref{fig-15}(a), the reconstructed images clearly illustrate lung impedance variations during respiration. As breathing proceeds, distinct impedance changes can be observed in different lung regions. During inhalation, the influx of air into the lungs reduces the local electrical conductivity, which is visualized as blue regions in the images. Conversely, during exhalation, the release of air increases the local conductivity, appearing as red regions. Fig.\ref{fig-15}(c) presents the GVBI variations corresponding to 100 frames of impedance data, and Fig.\ref{fig-15}(d) depicts the temporal variation of the summed pixel values within the ROI of the conductivity distribution images reconstructed from the same data set. A clear negative correlation is observed between GVBI and the ROI pixel-value sum. When the lungs are inflated with air, the GVBI reaches higher values, while the summed pixel values within the ROI of the reconstructed conductivity images decrease accordingly.

\section{Conclusion}

To overcome the limitations of conventional EIT systems, including high cost and bulky hardware, a high-precision, high-speed, low-cost, and portable impedance acquisition device is proposed in this work. The main research contents are as follows:
\begin{enumerate}
    \item An impedance acquisition device supporting two measurement modes, namely two-terminal serial and four-terminal parallel, was designed, enabling flexible mode switching to meet different measurement requirements. Moreover, voltage excitation is adopted in place of a current source, and the resulting excitation current is directly measured. This method avoids dynamic current adjustment and stringent DC offset suppression, leading to stable system performance. For enhanced portability and high-speed operation, multiplexers were adopted in place of relays for channel switching. To address the parasitic capacitance introduced by multiplexers, multiple optimization strategies were applied to suppress excitation and measurement oscillations and to minimize leakage currents in voltage sensing. For enhanced acquisition speed, five AD5933 devices are operated in parallel, with I²C bus management via the TCA9548 and a common external clock supplied by the ICS553 to ensure synchronized multi-channel measurements. Moreover, the device adopts lithium battery power and Bluetooth communication to improve portability.
    \item A PyQt5-based user-friendly GUI was developed to streamline system operation and enable efficient real-time imaging and data acquisition.
    \item A comprehensive performance evaluation of the electrical impedance tomography (EIT) system was performed. The results show that within the 8 kHz–100 kHz frequency range, the system achieves a stable signal-to-noise ratio of 50–70 dB, a relative standard deviation of less than 0.3\% across all frequencies, and a reciprocity error below 0.8\%, indicating robust measurement stability and reciprocity. In water-tank phantom experiments for small metal target imaging, the system successfully resolved metallic targets with diameters of 2.2–3.6 mm in the central region, with the ROI pixel-value sum increasing markedly as the target size increased. In addition, insulator imaging experiments yielded an average ICC of 83.25\%, demonstrating strong consistency between the reconstructed images and the ground-truth conductivity distribution. Furthermore, frequency-response experiments revealed that the imaging results vary with frequency in a manner consistent with the electrical characteristics of biological tissues. In lung respiration experiments, a pronounced correlation was observed between the reconstructed images and GVBI, enabling effective visualization of dynamic conductivity changes during respiration and validating the system’s applicability to real-world physiological imaging.
\end{enumerate}

\bibliographystyle{IEEEtran}
\bibliography{IEEEabrv, mybibfile.bib}

\end{document}